# Implementing Optimal Outcomes in Social Computing: A Game-Theoretic Approach

Arpita Ghosh

Yahoo! Research
Santa Clara, CA, USA
arpita@yahoo-inc.com

Patrick Hummel

Yahoo! Research
Berkeley, CA, USA
phummel@yahoo-inc.com

August 8, 2018

**Abstract**

In many social computing applications such as online Q&A forums, the best contribution for each task receives some high reward, while all remaining contributions receive an identical, lower reward irrespective of their actual qualities. Suppose a mechanism designer (site owner) wishes to optimize an objective that is some function of the number and qualities of received contributions. When potential contributors are *strategic* agents, who decide whether to contribute or not to selfishly maximize their own utilities, is such a "best contribution" mechanism, $\mathcal{M}_B$, adequate to *implement* an outcome that is optimal for the mechanism designer?

We first show that in settings where a contribution's value is determined primarily by an agent's expertise, and agents only strategically choose whether to contribute or not, contests *can* implement optimal outcomes: for any reasonable objective, the rewards for the best and remaining contributions in $\mathcal{M}_B$ can always be chosen so that the outcome in the unique symmetric equilibrium of $\mathcal{M}_B$ maximizes the mechanism designer's utility. We also show how the mechanism designer can learn these optimal rewards when she does not know the parameters of the agents' utilities, as might be the case in practice. We next consider settings where a contribution's value depends on both the contributor's expertise as well as her effort, and agents endogenously choose how much effort to exert in addition to deciding whether to contribute. Here, we show that optimal outcomes can never be implemented by contests if the system can rank the qualities of contributions perfectly. However, if there is *noise* in the contributions' rankings, then the mechanism designer can again induce agents to follow strategies that maximize his utility. Thus imperfect rankings can actually help achieve implementability of optimal outcomes when effort is endogenous and influences quality.

## 1 Introduction

Social computing systems where web users generate online content and create value in exchange for virutal rewards are now ubiquitous on the Web. In particular, there is an increasing number of online knowledge-sharing forums like Yahoo! Answers, Quora, StackOverflow, as well as non–English language forums like Naver (Korean) and Baidu Knows (Chinese), where users address or solve questions posed by other users. A number of these forums, such as Yahoo! Answers, MSN QnA, and Rediff Q&A, to name a few, are structured as a *contest*, where a contribution judged to be the best answer receives extra virtual points compared to the remaining answers, and users compete with each other to provide the best answer and collect these virtual points[1]. However, how *good* this best contribution is will depend on *which* users choose to enter and contribute an answer in response to the incentives provided by the system, since entry is endogenous, *i.e.*, a user may choose to simply not participate in the system. In addition, viewers may also derive value from contributions other than the one judged to be best answer— for instance, in Q&A sites like Yahoo! Answers, there are questions (such as what to do over a weekend in New York) that may not have a single objective answer, and more than one answer may provide value.

Is a contest the 'right' structure for eliciting desirable outcomes in such Q&A forums, in the presence of strategic participants acting selfishly to maximize their own utility? In this paper, we address this question from a game-theoretic perspective. Users with varying abilities have a cost to answer a question or complete a task, and the quality, or value, of their output, is a function of their ability and possibly the effort they expend. A mechanism in this setting specifies how to reward agents for their contributions: for example, a contest or "best contribution" mechanism would assign some high reward of $p_B$ points to the winner and a lower reward of $p_C \in [0, p_B)$ points to all other contributors, while an alternative mechanism may split a total prize $P$ amongst participants in proportion to the quality of their contributions.

In such a scenario, the realized outcome— the number and qualities of elicited contributions— depends on the strategic choices of agents responding to the incentives provided by the mechanism. Suppose the mechanism designer (*i.e.*, the owner of the site) can quantify the desirability of each such outcome, via some function of the number and qualities of the elicited contributions. Is it possible, for *any* choice of mechanism at all, to support an outcome that is optimal for the mechanism designer *in an equilibrium* of the corresponding game? And more specifically, is the simple best contribution mechanism, $\mathcal{M}_B$, with only *two* parameters $p_B$ and $p_C$ to vary, adequately powerful to implement an optimal outcome?

**Our Contributions.** We present a model (§2) to address the question of whether contests, modeled as best contribution mechanisms $\mathcal{M}_B$, are adequate to *implement* optimal outcomes in the presence of strategic contributors. We address this question for two types of situations that arise in online Q&A forums, depending on whether the value of a contribution comes primarily from the contributors' *ability*, as in expertise-based questions, or whether it is a function both of the agent's ability and her endogenously chosen effort, as in research-intensive or thought-intensive questions. Our results hold for a fairly general class of utility functions $V$ of the mechanism designer, as well as with general amounts of noise or error in the rankings of the contributions by the mechanism: we only require that $V$ is increasing in quality for any *fixed* number of contributions, and that the ranking function is such that an agent's probability of winning increases with the quality of her contribution.

We first consider a model where agents with heterogeneous abilities strategically choose only whether or not to contribute, modeling settings where an agent's expertise rather than effort primarily determines the value of her contribution. Our main result for this model shows that surprisingly, best contribution mechanisms are indeed adequately powerful to implement optimal outcomes for a broad class of objectives $V$— the rewards for the winning and remaining contributions can always be chosen so that contributors follow strategies maximizing the mechanism designer's utility in the unique symmetric equilibrium. The values of these optimal rewards can be learned by the mecha-nism designer when she does not know the parameters of the agents' utilities, as might be the case in practice. We next address the question of whether there exist asymmetric equilibria, and prove by construction that suboptimal asymmetric equilibria can indeed exist. However, this negative result is not limited to best contribution mechanisms, and can be extended to a more general class of mechanisms— the optimal outcome need not be the unique equilibrium for general rank-order mechanisms if asymmetric equilibria are considered as well.

---
[1] Virtual rewards indeed seem to be a strong incentive for users of online systems: [13], [15], and [16] all present evidence that virtual points tend to motivate contributions from users.

We then consider endogenous effort settings where the value of an agent's contribution is a function both of her ability and effort, and agents strategically choose both whether or not to participate as well as how much effort they expend on their contributions. We first show that in this setting, optimal outcomes cannot be implemented if the system ranks the qualities of the contributions perfectly— agents never all follow strategies that maximize the mechanism designer's utility in any equilibrium of the game. However, these adverse incentives can be avoided if the system does not rank contributions perfectly— the mechanism designer may then again be able to choose the rewards in the best contribution mechanism so as to implement the optimal outcome. Thus when effort is endogenous and influences quality, it can actually be beneficial for the mechanism designer to use noisy rankings in order to achieve implementability of optimal outcomes.

## 1.1 Related Work

There has been some prior work on the design of incentives for online Q&A forums [6], addressing the issue of delayed responses. [6] studies the problem of designing incentives for users to contribute their answers quickly, rather than waiting to make their contributions; it does not distinguish between the qualities of responses provided by different participants. Our work relates more closely to that on the optimal design of crowdsourcing contests ([1], [5], [8]), as well as a large body of work in the economics literature on the design of contests to optimally incentivize participants ([3], [7], [9], [10], [11], [12], [14]). The key differences between all these papers and our work are the following. First, we consider an environment with virtual points which are, to a first-order approximation, costless to the mechanism designer, whereas these papers all consider settings where the mechanism designer pays real money to agents, and consequently the size of these payments has a direct effect on the mechanism designer's utility. Second, this literature addresses the question of what are the best outcomes that can be supported in equilibrium, *i.e.*, the question of optimizing *amongst implementable outcomes*, rather than *whether* the optimal outcome, assuming nonstrategic agents, is implementable by any mechanism in the presence of strategic agents. Finally, we study a model with endogenous entry, *i.e.*, agents can strategically decide whether or not to participate, and also, we allow the mechanism designer's utility to depend in a general way on number and qualities of contributions, in contrast to specific objectives like maximizing the highest effort or the sum of the top $k$ efforts for some fixed $k$ (see §2 for a motivation of more general utilities).

The work most closely related to ours from the crowdsourcing contest design literature is that of [1], which considers a game-theoretic model of a crowdsourcing contest and asks how to optimally split a prize budget amongst contestants to achieve the maximum equilibrium effort. In contrast, we ask whether a mechanism designer can achieve an equilibrium outcome with utility equal to the maximum possible utility achievable with nonstrategic agents, rather than how to optimize over the set of implementable outcomes. [5] models crowdsourcing as an all-pay auction, where all agents pay a cost to exert effort, but only one winner obtains the benefit from winning the auction, and addresses the question of how to design an all-pay auction to maximize the highest effort, in contrast with the conventional objective of maximizing revenue, or total effort. In contrast, we ask whether the family of contest-style mechanisms is powerful enough to optimize the mechanism designer's objective and how payoffs must be structured to incentivize this desired optimal behavior.

The economics literature on how firms can use contests to create incentives for employees to work hard ([3], [7], [9], [10], [12]) again addresses questions related to using real money to optimize amongst a set of implementable outcomes, and does not address whether a mechanism designer can achieve her most preferred outcome over all possible outcomes in a setting with virtual rewards. Also, this body of work typically assumes a specific functional form for the mechanism designer's utility, and does not address how endogenous participation may affect this utility.

## 2 Model

There are $n$ potential contributors, or agents, $i = 1, \ldots, n$, who strategically make contribution choices in a single question or task in an online Q&A forum. Different agents have different abilities, or levels of expertise for any given question or task. We denote the ability of an agent $i$ as $a_i$; an agent who is more capable has a higher value of $a_i$ than a less capable agent.

The abilities $a_i$ are independent and identically distributed draws from a distribution with CDF $F$. We assume that $F$ is atomless and strictly increasing on its support, which we assume without loss of generality is the bounded interval $[0, 1]$. An agent knows her own ability $a_i$ but not the abilities $a_j$ drawn by the remaining agents; the distribution $F$, however, is common knowledge to all parties.

The quality of the contribution produced by an agent $i$, which we denote by $q_i$, is a non-decreasing function of her ability $a_i$, as well as possibly the *effort $e_i$* she puts in, *i.e.*, $q_i = q_i(a_i, e_i)$.

We model two different kinds of situations that arise in online Q&A forums. In some settings such as knowledge-based Q&A forums, agents possess differing amounts of knowledge about a question and further knowledge is not easily obtainable— here the quality, or the value provided by the contribution is determined largely by the expertise of the contributor rather than her effort, and the cost to answering is only the effort of transcribing this knowledge. However, there are also settings where a contributor might be able to significantly increase the quality or value of her contribution with additional effort, for instance on questions on sites like StackOverflow which might request code to accomplish some task. Here, the quality of a contribution depends both on the contributor's expertise and effort. We model these two different settings via the homogenous and endogenous effort models below.

1. *Homogenous effort:* In the homogenous effort model, every agent incurs the same effort $e$ and pays the same cost $c_C$ if she chooses to make a contribution. The differences in the realized qualities of contributions then arise solely from the differences in agents' abilities. Without loss of generality, we may write $q_i(a_i, e_i) = a_i$.

   The homogenous effort model corresponds to a situation in which agents differ in the extent to which they are able to answer a question, but it would take all individuals similar levels of effort to actually contribute their knowledge as, for example, in the context of a question on health in Yahoo! Answers.

2. *Endogenous effort:* In the endogenous effort model, each agent $i$ can endogenously choose to exert some level of effort $e_i \in [0, 1]$ if she decides to contribute. An agent who exerts effort $e$ pays a cost $c(e)$, where $c(e)$ is a continuously differentiable and increasing function of

e. The final quality of an agent's contribution, $q_i(a_i, e_i)$, is a continuously differentiable and increasing function of the agent's effort $e_i$ and her ability $a_i$.

Note that agents are heterogeneous in both these models, as captured by their different abilities $a_i$.

*Actions.* Agents in our model can strategically choose whether or not to contribute, as well as how much effort to put in conditional on contributing in the endogenous effort model. In addition, to fully capture the set of actions available to agents for gaining points on real Q&A forums such as Y! Answers, we also allow agents to have the option of *rating* the received contributions: an agent who decides to rate incurs a cost of $c_R \geq 0$[2]. We emphasize, however, that all our results *continue to hold* if agents do not have the option to rate, or do not receive points for rating.

Formally, each agent $i$ strategically chooses an action $\alpha_i \in \{C, R, N\}$, where $\alpha_i = C$ indicates that agent $i$ chooses to *contribute*, $\alpha_i = R$ indicates that agent $i$ *rates* (some nonempty subset of) the received contributions, and $\alpha_i = N$ indicates that agent $i$ does not participate at all[3]. In the endogenous effort model, an agent who chooses to contribute (*i.e.*, $\alpha_i = C$) can additionally choose her level of effort $e_i \in [0, 1]$.

After agents strategically choose their actions, a *mechanism* allocates rewards in the form of points to the agents. We denote the number of agents that decide to contribute by $m$. Let $\alpha \equiv (\alpha_1, \ldots, \alpha_n)$ denote the vector of action choices of the agents and $q \equiv (q_1, \ldots, q_m)$ denote the vector of the qualities of received contributions. In general, an agent's expected reward can depend on her choice of action, the quality of her output if contributing, the actions of the other agents, and the qualities and number of contributions produced by other agents. Let $p_i(\alpha, q)$ denote the expected number of points awarded to agent $i$ by the mechanism. An agent's expected payoff for the game is equal to the expected number of points she obtains minus any costs associated with her choice of action:
$$u_i = p_i(\alpha, q) - c(\alpha_i),$$
where $c(\alpha_i)$ denotes the cost agent $i$ incurs from taking action $\alpha_i$. An agent who does not participate, *i.e.*, chooses $\alpha_i = N$, always receives no benefit and pays no cost, for a payoff of 0.

**Mechanisms.** There are many different ways to reward agents as a function of their actions. For instance, a rank-order mechanism awards points to an agent based on how highly her contribution ranks amongst all contributions (rather than its absolute quality), while a proportional mechanism would award points in proportion to the realized qualities of the contributions. In this paper, we will focus on the family of *best contribution* mechanisms $\mathcal{M}_B$, which choose a best or 'winning' contribution and give the winner a higher reward than the other contributors. This mechanism is of particular interest because it is widely used in many settings where extra points are awarded for the 'best' user-generated contribution such as in MSN QnA, Rediff Q&A, and Yahoo! Answers, as well as in crowdsourcing contests where the best submission receives a prize or award (and the remaining submissions usually receive nothing).

**Definition 2.1.** *The best contribution mechanism $\mathcal{M}_B(p_B, p_C)$ selects some contribution as the*

---
[2] Note that this cost $c_R$ is the total cost for rating, not the cost of rating an individual contribution. This is reasonable because we do not reward agents for individual ratings either—agents are not required to rate all contributions, nor to rate with high accuracy, to earn the reward for rating

[3] Since agents would not have an incentive to rate honestly if they were allowed to rate their own contributions, we assume that an agent can choose to either contribute or rate contributions for a given question but *not both*; this is in keeping with the practice on several websites such as Slashdot.

*best or winning contribution, and awards the corresponding contributor or winner $p_B > 0$ points and all remaining contributors $p_C \in [0, p_B)$ points.*

An agent who rates (any nonzero number of) contributions receives $p_R \geq c_R$ points, but an agent receives no payoff from choosing the rating action if there are no contributions to rate. The assumption that $p_R \geq c_R$ ensures that agents prefer to rate than to not participate at all. We note here that a system which does not allow or reward rating is easily captured in our model by setting $p_R = 0$; identical results can be obtained with substantively identical proofs in this case where $p_R \leq c_R$ and all agents weakly prefer not to participate than to rate.

Given a set of contributions, the best contribution mechanism $\mathcal{M}_B$ needs to pick a winning contribution. We use a function $\pi$ to describe how the mechanism $\mathcal{M}_B$ chooses a winner as a function of the qualities of received contributions. Note that we do not require the mechanism $\mathcal{M}_B$ to be able to identify the highest quality contribution with perfect accuracy (in fact, as we will see in §4, this might not even be desirable). Instead our results only require that higher quality contributions are more likely to be chosen as winners.

**Definition 2.2.** *We use $\pi(q_i, q_{-i}, m)$ to denote the probability that an agent $i$ with a contribution of quality $q_i$ will be chosen as the winner when there are $m$ other agents who produce contributions with qualities $q_{-i}$.*

We make the following assumptions on $\pi$:

1. $\pi(q_i, q_{-i}, m)$ is increasing in $q_i$ and decreasing in $m$. That is, agent $i$ should be more likely to win if she makes a higher quality contribution, and less likely to win if the number of other contributors increases.

2. For any fixed values of $m$ and $q_i^*$, the set of values of $q_{-i}$ for which $\pi(q_i, q_{-i}, m)$ is discontinuous in $q_i$ at $q_i = q_i^*$ has measure zero.

Note that the ratings from the agents who choose to rate may or may not be used towards determining the best contribution: as long as the agents rating content are more likely to rate a higher quality contribution more highly, the probability $\pi(q_i, q_{-i}, m)$ is likely to satisfy these conditions.

**Solution Concept.** Throughout this paper, we use the solution concept of a Bayes-Nash equilibrium, in which each player takes an action that maximizes her expected utility given the strategies of the other agents and the unknown abilities of the other agents. This an appropriate solution concept when an agent knows her own ability but not that of the other agents when taking an action, as in our setting.

**Mechanism Designer's Objective.** We suppose that the mechanism designer can quantify the desirability of each possible outcome, specified by the number of contributors $m$ and their qualities, via some function $V(m, q^1, \ldots, q^m)$, where $q^i$ denotes the quality of the $i^{th}$ best contribution. We will assume that the function $V$ is continuous in $(q^1, \ldots, q^m)$ for all $m$, and that it is nondecreasing in $\{q\}$ for a given $m$. Such a utility function $V$ is fairly general, capturing a wide variety of possible preferences that a mechanism designer might have— the only requirement we impose on $V$ is that, for any *fixed* number of contributions $m$, the mechanism designer must prefer higher quality contributions to lower quality ones.

Specifically, our condition on $V$ allows for *non-monotonicity* in the *number* of contributions— a larger number of contributions may not always be more valuable, so that, for example, the outcome $(q_1, q_2)$, might be preferred to the outcome $(q_1, q_2, q_3)$ even for large values of $q_3$. Accommodating such nonmonotonicity is important because it allows modeling situations where there is a cost to *searching* for higher quality contributions, so that the mechanism designer may actually prefer to limit the number of contributions to only those with adequately high quality. In addition, it also allows capturing situations where a mechanism designer may value having multiple high quality answers much more than he values each answer individually, such as when these answers reinforce each other and increase the asker's confidence that he is getting high quality advice that he can act on. Such a utility function, where for example $V(2, q, q') > V(1, q) + V(1, q')$, is not captured by the standard objectives of highest or total quality.

## 3 Homogenous Effort

We begin with addressing the question of whether the parameters of the best contribution mechanism $\mathcal{M}_B$ can always be chosen so as to implement optimal outcomes in the homogenous effort model. In addition to capturing expertise-based settings where effort does not significantly affect the value of a contribution, the equilibrium analysis in this simpler model serves as a useful building block for the analysis in the endogenous effort model.

We first analyze the equilibria of the family of mechanisms $\mathcal{M}_B$ in §3.1, and show that there is a unique symmetric equilibrium in threshold strategies, where this threshold ability is a continuous, monotone function of the rewards $(p_B, p_C)$. Next we prove that as long as the mechanism designer's utility increases with quality (although not necessarily with participation), some symmetric threshold strategy maximizes her utility as well. This, together with the continuity and monotonicity properties of the equilibrium threshold, can be used to show that there exists a choice of rewards that implements the optimal outcome (§3.2).

### 3.1 Equilibrium Analysis

We begin with the existence of a symmetric threshold strategy equilibrium.

**Lemma 3.1** (Existence). *For any values of $p_B$ and $p_C$ with $p_B > p_C$, there exists a symmetric equilibrium in* threshold strategies *for the mechanism $\mathcal{M}_B(p_B, p_C)$: there is a threshold $a^*(p_B, p_C)$ such that it is an equilibrium for each agent $i$ to contribute if and only if her ability $a_i \geq a^*(p_B, p_C)$.*

*Proof.* First note that regardless of the strategies of the remaining agents, agent $i$ has a unique best response, which is a threshold strategy $a_i^*$, where $i$ contributes (*i.e.*, chooses action $C$) if and only if her ability $a_i \geq a_i^*$. To see this, note that agent $i$'s expected payoff from rating is independent of her ability $a_i$, but her payoff from contributing is strictly increasing in her ability $a_i$ because her probability of winning is strictly increasing in $a_i$. Thus if agent $i$ prefers contributing to rating at ability $a_i$, she must strictly prefer contributing to rating at all abilities $a_i' > a_i$. Similarly, if she prefers rating to contributing at ability $a_i$, she must also strictly prefer rating to contributing at all abilities $a_i' < a_i$. Thus regardless of the strategies of the other agents, an agent $i$'s unique best response is to use a threshold strategy $a_i^*$ in which the agent contributes if and only if $a_i \geq a_i^*$.

It follows that an equilibrium in the related game in which agents are restricted to using threshold strategies is also an equilibrium of the original game. But the game in which agents are restricted to using threshold strategies is a symmetric game in which each agent has a set of possible actions that is compact and Hausdorff and each agent's expected utility is a continuous function of the actions of the agents. Thus by Theorem 1 of [4], it follows that there exists a symmetric mixed strategy equilibrium in the game in which agents are restricted to using threshold strategies. Now since an agent has a unique best response to any strategy choices of the remaining agents, it follows that the agents must not be randomizing over different thresholds in this equilibrium. Thus there exists a symmetric threshold equilibrium in which all agents use the same threshold $a^*$. □

For the remaining results in this section it will be useful to introduce the following notation.

**Definition 3.1.** *[$Pr(\mathcal{C}_{>0}|a,n), Pr(\mathcal{W}|a,n)$] Recall that agents' abilities $q_i$ are drawn from the distribution $F$. We let $Pr(\mathcal{C}_{>0}|a,n)$ denote the probability that an agent sees at least one contribution from the other agents when these remaining $n-1$ agents each use a threshold strategy with threshold $a$. We let $Pr(\mathcal{W}|a,n)$ denote the probability that an agent with ability $a$ 'wins', i.e., is chosen as the best contribution, when the remaining $n-1$ agents are using a threshold strategy with the same threshold $a$. We drop the dependence on $n$ when the value of $n$ is fixed and clear from the context.*

Note that $Pr(\mathcal{C}_{>0}|a,n) = 1 - F^{n-1}(a)$. If contributions are perfectly ranked according to their qualities, then $Pr(\mathcal{W}|a,n) = F^{n-1}(a)$, since the only way an agent with ability $a_i = a$ can win against other agents $j$ who contribute according to the threshold $a$ is if none of them contributes. However, since we allow for the possibility that contributions are not ranked perfectly according to their qualities, in general $Pr(\mathcal{W}|a,n)$ may be greater than $F^{n-1}(a)$ under our model.

Now we prove that there is a unique symmetric equilibrium to $\mathcal{M}_B(p_B, p_C)$.

**Theorem 3.1** (Uniqueness). *The mechanism $\mathcal{M}_B(p_B, p_C)$ has a unique symmetric equilibrium.*

*Proof.* Since any individually rational strategy is a threshold strategy, any equilibrium must be in threshold strategies. Thus it suffices to show that there is a unique symmetric threshold strategy equilibrium.

First note that since $F$ is an atomless distribution, the probability $Pr(\mathcal{C}_{>0}|a^*)$ that another agent contributes is strictly decreasing and continuous in $a^*$, and similarly $Pr(\mathcal{W}|a^*)$ is strictly increasing in $a^*$ (because the agent faces less competition in the sense of first order stochastic dominance when $a^*$ increases) and continuous in $a^*$ (because the random amount of competition the agent faces varies continuously with $a^*$).

Therefore, the expected utility an agent obtains from choosing to rate when all agents play according to a threshold strategy $a^*$,

$$u^R(a^*) = (p_R - c_R)Pr(\mathcal{C}_{>0}|a^*),$$

is nonincreasing and continuous in $a^*$, and the expected utility an agent $i$ with ability $a_i = a^*$ obtains from contributing when all other agents are using a threshold strategy with threshold $a^*$,

$$u^C(a^*) = (p_B - c_C)Pr(\mathcal{W}|a^*) + (p_C - c_C)[1 - Pr(\mathcal{W}|a^*)],$$

is strictly increasing and continuous in $a^*$.

Now if there is a value $a^*$ such that an agent with ability $a_i = a^*$ is indifferent between rating and contributing when all other agents play according to the threshold $a^*$, i.e., if there is a solution $a^* \in [0,1]$ to the equation

$$u^R(a^*) = u^C(a^*),$$

then this threshold $a^*$ constitutes an equilibrium. Since $u^R$ is nonincreasing in $a^*$ and $u^C$ is strictly increasing in $a^*$, such a solution $a^*$, if it exists, is unique, and gives the unique symmetric equilibrium to $\mathcal{M}_B(p_B, p_C)$.

If there is no such $a^*$, then either $u^R(a^*) > u^C(a^*)$ for all $a^* \in [0,1]$ or $u^R(a^*) < u^C(a^*)$ for all $a^* \in [0,1]$. In the former case, the only equilibrium is for all agents to always rate (corresponding to the threshold $a^* = 1$), and in the latter case the only equilibrium is for all agents to always contribute (corresponding to the threshold $a^* = 0$). Therefore, in all of these cases there exists a unique symmetric threshold equilibrium. □

We note that if $p_C - c_C > p_R - c_R$, then an agent always prefers contributing to rating even if she knows that she will not have the best contribution. Similarly, if $p_B - c_C < 0$, an agent always prefers not to contribute even if she knows that she will have the best contribution. In these cases, the symmetric threshold equilibrium will have thresholds $a^* = 0$ and $a^* = 1$ respectively. But in intermediate cases, where
$$p_C - c_C < p_R - c_R < p_B - c_C,$$
the symmetric threshold equilibrium must have $a^* \in (0,1)$, so that agents neither always rate nor always contribute.

Next we prove that the unique equilibrium threshold $a^*(p_B, p_C)$ varies continuously with $p_B$ and $p_C$, which we use to prove the main implementation result in the next section. This result guarantees that agents will not dramatically change the strategies they use as a result of small changes in the rewards that the mechanism designer is using to incentivize the agents.

**Theorem 3.2** (Continuity). *The equilibrium threshold $a^*(p_B, p_C)$ is continuous in $p_B$ and $p_C$.*

*Proof.* The equilibrium threshold $a^*$ is given by the unique solution, if it exists, to

$$u_\Delta(p_B, p_C, a) \equiv (p_R - c_R)Pr(\mathcal{C}_{>0}|a) - (p_B - c_C)Pr(\mathcal{W}|a) - (p_C - c_C)[1 - Pr(\mathcal{W}|a)] = 0,$$

where $Pr(\mathcal{C}_{>0}|a)$ and $Pr(\mathcal{W}|a)$ are as in Definition 3.1.

Note that if $p_B$ and $p_C$ change by an amount no greater than $\delta > 0$, then $u_\Delta(p_B, p_C, a)$ changes by an amount $O(\delta)$. If $a$ increases (decreases) by $\epsilon$, then $Pr(\mathcal{C}_{>0}|a)$ decreases (increases) by $\Omega(\epsilon^{n-1})$ and $Pr(\mathcal{W}|a)$ increases (decreases) by $\Omega(\epsilon^{n-1})$, meaning $u_\Delta(p_B, p_C, a)$ decreases (increases) by $\Omega(\epsilon^{n-1})$. Thus for any $\epsilon > 0$ there exists some $\delta > 0$ such that if $p_B$ and $p_C$ change by an amount no greater than $\delta$, then the value of $a$ that satisfies $u_\Delta(p_B, p_C, a) = 0$ changes by an amount no greater than $\epsilon$. Thus the equilibrium threshold is continuous in $p_B$ and $p_C$ if there exists some $a$ satisfying $u_\Delta(p_B, p_C, a) = 0$.

If there is no $a$ satisfying $u_\Delta(p_B, p_C, a) = 0$, then either all agents strictly prefer to rate when all other agents are rating or all agents strictly prefer to contribute when all other agents are contributing. Small changes in $p_B$ and $p_C$ would not affect these strict preferences, so it would remain an equilibrium for all agents to rate or for all agents to contribute after small changes in $p_B$ and $p_C$. Thus the equilibrium threshold is also continuous in $p_B$ and $p_C$ when there does not exist a $a$ satisfying $u_\Delta(p_B, p_C, a) = 0$. □

## 3.2 Implementability

Having characterized equilibrium strategies in the mechanism $\mathcal{M}_B$ in the previous section, we now prove the main implementation result for the homogenous effort model. We show that, regardless of the precise form of the mechanism designer's utility function $V(m, q^1, \ldots, q^m)$, the values of the rewards $p_B$ and $p_C$ in the best contribution mechanism $\mathcal{M}_B$ can always be chosen to induce agents to follow strategies that maximize the mechanism designer's expected utility.

The proof of this implementation result breaks down into two parts. First we show that for any utility function $V$, $E[V]$ can be maximized by a threshold strategy $\hat{a}$ in which each agent $i$ chooses to contribute if and only if $a_i \geq \hat{a}$. We then show that for any $\hat{a}$, $p_B$ and $p_C$ can always be chosen such that agents use the threshold strategy $\hat{a}$ in equilibrium in $\mathcal{M}_B$.

Throughout this section, we will restrict attention to symmetric strategies: specifically, the strategy maximizing the mechanism designer's expected utility is the optimal strategy *amongst* symmetric strategies.

**Lemma 3.2.** *There exists a threshold strategy, where each agent contributes if and only if her quality is greater than a common threshold $\hat{a}$, that maximizes the mechanism designer's expected utility.*

*Proof.* Suppose there exists some strategy $\sigma$ such that the mechanism designer's expected utility $E[V]$ is maximized when all agents use the strategy $\sigma$. Let $\lambda(\sigma)$ denote the probability (over random draws of agents' abilities from $F$ and any randomness in $\sigma$) with which an agent contributes when using the strategy $\sigma$. Consider the alternative threshold strategy $\hat{a}(\sigma)$ whereby an agent $i$ contributes if and only if her ability $a_i \geq \hat{a}(\sigma)$, where $\hat{a}(\sigma) = F^{-1}(1 - \lambda(\sigma))$ is chosen so that the probability an agent contributes under this threshold strategy remains unchanged at $\lambda(\sigma)$. Such a $\hat{a}$ always exists because $F(a)$ is continuous in $a$ on its support.

Now the mechanism designer's expected utility $E[V]$ is at least as large when the agents use this threshold strategy as it is when they use the strategy $\sigma$: to see this, note that the distribution of the number of agents who contribute is the same regardless of whether agents use the strategy $\sigma$ or the threshold $\hat{a}(\sigma)$. But conditional on contributing, an agent's distribution of qualities under the threshold strategy $\hat{a}(\sigma)$ first order stochastically dominates this distribution under the strategy $\sigma$. Since $V$ is increasing in the quality of the contributions by assumption, the mechanism designer's expected utility when agents are using the threshold strategy $\hat{a}(\sigma)$ is at least as large as that when agents are using the strategy $\sigma$.

Therefore, there exists a (symmetric) strategy that maximizes $E[V]$ if and only if there exists a threshold $\hat{a}$ that maximizes $E[V]$ amongst the class of (symmetric) threshold strategies. But such an optimal threshold always exists because $E[V]$ is continuous in $\hat{a}$ and there is a compact set of possible thresholds $\hat{a}$. The result then follows. $\square$

From Lemma 3.1 and Theorem 3.1, we know that agents use the threshold strategy $a^*$ in equilibrium, and from Lemma 3.2, we know that the mechanism designer's utility function is maximized when agents use the threshold strategy $\hat{a}$. So to prove the implementation result, we only need to show that these thresholds can be made to coincide. We now show that for any $\hat{a} \in [0,1]$, there exists a choice of $p_B$ and $p_C$ such that the threshold $a^*(p_B, p_C)$ that agents use in equilibrium in the mechanism $\mathcal{M}_B(p_B, p_C)$ satisfies $a^*(p_B, p_C) = \hat{a}$.

**Theorem 3.3.** *For any $p_R$ and any fixed ratio of the rewards $p_B/p_C > 1$, there exist values of $p_B$ and $p_C$ such that the unique symmetric threshold equilibrium of $\mathcal{M}_B(p_B, p_C)$ maximizes the mechanism designer's expected utility.*

*Proof.* Note that if $p_B - c_C < 0$ and $p_C - c_C < 0$, then the unique equilibrium is for all agents to rate, *i.e.*, $a^* = 1$. If $p_B - c_C > p_C - c_C > p_R - c_R$, then the unique equilibrium is for all agents to contribute, corresponding to the threshold $a^* = 0$. Since $a^*$ is continuous in $p_B$ and $p_C$ (by Lemma 3.2), it follows from the intermediate value theorem that there exist values of $p_B$ and $p_C$ such that $a^*(p_B, p_C) = \hat{a}$ for any value of $\hat{a} \in [0, 1]$. And since we know from Lemma 3.2 that the mechanism designer's expected utility can always be optimized by using a threshold strategy, the result follows. □

This result indicates that the commonly used best contribution mechanism is actually quite powerful for inducing agents to follow strategies that are optimal for the mechanism designer when the main difference between agents is how they differ in their levels of expertise. We note the generality of this result: (i) it holds for arbitrary utility functions of the mechanism designer (provided they depend only on the number and qualities of contributions) and (ii) it holds even with general amounts of noise or errors in the rankings of the contributions; *i.e.*, the probability a contribution may be ranked best even if it is not the highest quality contribution can depend in a very general way on the number and qualities of the contributions. Our implementation result in this section continues to hold as long as $V$ is increasing in quality and an agent's probability of winning increases with quality (*i.e.*, higher quality contributions are more likely to win).

Next we illustrate how the mechanism designer would want to choose the values of $p_B$ and $p_C$ in practice. First we show how the values of $p_B$ and $p_C$ that must be used to induce the optimal outcome vary as a function of the number of agents, $n$. This is relevant because some types of questions differ systematically in the numbers of agents that arrive. The following result is proven and discussed in the appendix.

**Proposition 3.1** (Comparative Statics). *Suppose that the threshold the mechanism designer would like the agents to use in equilibrium is independent of $n$. For any fixed $p_R$ and any fixed ratio $p_B/p_C > 1$, the numbers of points the mechanism designer should award as a function of $n$, $p_B(n)$ and $p_C(n)$, are strictly increasing in $n$.*

Next we address how the mechanism designer can learn the values of agents' utility functions when the mechanism designer does not initially know how costly it is for agents to make contributions. In order to compute the values of the rewards $p_B$ and $p_C$ that implement optimal outcomes in $\mathcal{M}_B$, the mechanism designer needs to know the values of the agents' costs $c_C$ for making a contribution, but these may not be known in practice. The following result, proven and discussed in the appendix, shows how the mechanism designer can conduct a series of contests to learn the cost of contributing if the cost of rating, $c_R$, is known (a plausible assumption since this cost is often close to zero).

**Theorem 3.4.** *Suppose the mechanism designer conducts a series of $T$ contests with $n$ agents in which the mechanism designer awards $p_B > p_R + \overline{c}$ points for winning and $p_C = 0$ points for contributing but not winning, and let $n_C^t$ denote the number of agents who choose to contribute in the $t^{th}$ contest. Then, $\sum_{t=1}^{T} n_C^t/(Tn)$ converges in probability to an invertible function of $c_C$ as $T \to \infty$.*

When $c_R$ is unknown, the mechanism designer can learn the values of $c_C$ and $c_R$ by conducting two separate series of contests of the form in Theorem 3.4; we discuss the extension of this result to cases where $c_R$ is unknown in the appendix.

**Implementation for General Mechanisms** We have seen that the mechanism $\mathcal{M}_B$, which awards a strictly larger number of points to a single best contribution, and a flat number of points to all remaining contributions, is a rather powerful mechanism in terms of creating optimal incentives for agents. Is $\mathcal{M}_B$ special in this regard, or does this implementation result hold for other mechanisms as well? In fact, this result does hold for other mechanisms— the key property that is needed for the implementation result is the monotonicity and continuity of an agent's expected payoff from contributing as a function of quality.

Consider a more general class of mechanisms where the expected reward an agent with quality $q_i$ obtains from contributing when $m$ other agents contribute with qualities $q_{-i}$ is some function $p(q_i, q_{-i}, m)$. Suppose that $p(q_i, q_{-i}, m)$ is strictly increasing in $q_i$ and continuous in $(q_i, q_{-i})$. The best contribution mechanism $\mathcal{M}_B$ that we study is a special case of this general class of mechanisms, but there are many other mechanisms with this property. For example, a mechanism in which the highest ranked contribution receives $p_1$ points, the contribution ranked second receives $p_2 < p_1$ points and so on, satisfies these criteria as long as the probability that agent $i$ is ranked ahead of agent $j$ is increasing in $q_i$ and continuous in $q_i$ and $q_j$. Another example is a proportional mechanism in which $p(q_i, q_{-i}, m) = \frac{q_i}{\sum_j q_j}$ (with appropriate values when $q_i = 0$ for all $i$).

The implementation result in fact continues to hold for this more general class of mechanisms as well. The proof is very similar to the argument for the best contribution mechanism $\mathcal{M}_B$ in §3.1 and 3.2. Lemma 3.1, which guarantees existence of a symmetric threshold equilibrium, only relies on the fact that an agent is more willing to contribute when she has a greater ability, which remains true for this more general class of mechanisms, since an agent obtains more points in expectation from contributing with greater ability. The continuity of the equilibrium threshold in the number of points awarded, given in Theorem 3.2, will continue to hold because $p(q_i, q_{-i}, m)$ is continuous in $(q_i, q_{-i})$.

Given this, if the mechanism designer awards an expected number of points equal to $kp(q_i, q_{-i}, m)$ for some $k > 0$ chosen by the mechanism designer, $k$ can be chosen to induce the agents to follow a strategy that maximizes the mechanism designer's expected utility: when $k$ is very small, then agents will never want to contribute, whereas when $k$ is very large, agents will always want to contribute. From the continuity of strategies in points, it then follows that by appropriately choosing an intermediate value of $k$, the mechanism designer can induce any intermediate rates of contribution characterized by any threshold $\hat{q} \in (0, 1)$.

## 3.3 Asymmetric Equilibria

The implementation result we proved in §3.2 guarantees uniqueness of the optimal outcome amongst the set of *symmetric* equilibria— there exists a choice of $p_B$ and $p_C$ such that the *unique symmetric* equilibrium of $\mathcal{M}_B(p_B, p_C)$ consists of strategies that maximize the mechanism designer's utility. But what about asymmetric equilibria— can there exist other, asymmetric, equilibria in $\mathcal{M}_B$?

In this section, we show that the answer to this question, unfortunately, is yes. However, as we show, this is not a difficulty only with the mechanism $\mathcal{M}_B$— asymmetric equilibria can arise in a more general class of mechanisms as well, even with a general amount of noise or error in the rankings of the contributions. This indicates that, even for fairly general classes of mechanisms, it is not possible to guarantee that the equilibrium maximizing $E[V]$ will be the unique equilibrium of the game, if we also consider asymmetric equilibria.

We consider a general class of mechanisms in which an agent ranked first receives $p_1$ points, an agent ranked second receives $p_2$ points, and in general the $k$-th ranked agent receives $p_k$ points. These values of $p_k$ do not depend directly on the precise quality of any agent's contribution but instead depend only on the ordering of the contributions. We assume throughout that $p_1 \geq p_2 \geq \ldots \geq p_n$ and at least one of these inequalities is strict.

In order to allow for a wide range of accuracies with which contributions are ranked, we introduce a parameter $\beta \in [0, 1]$ which represents the extent to which the system is able to distinguish contributions on the basis of their qualities. With probability $\beta$, the system is able to perfectly distinguish the qualities of the contributions and ranks a contribution $i$ ahead of another contribution $j$ if and only if $q_i > q_j$. With probability $1 - \beta$, the system is unable to distinguish the qualities of any of the contributions, and simply orders them randomly. Larger values of $\beta$ indicate that it is more likely that the system will accurately order the contributions by quality. While this is not the most general possible formulation, it suffices to illustrate our point.

Under this general class of mechanisms and general levels of system accuracy in ranking contributions, there always exists the possibility of asymmetric equilibria, as shown by the following result.

**Theorem 3.5.** *For any values of $p_k$ and $\beta$, there exists an asymmetric equilibrium as long as it is not an equilibrium for all agents to rate or for all agents to contribute.*

*Proof.* We construct an equilibrium in which agent 1 always contributes (corresponding to the threshold $a_1^* = 0$) and the remaining agents $i = 2, \ldots, n$ contribute according to a threshold $a_i^* = a^* > 0$. Note that by the assumptions on the distribution $F$, $F(a^*) > 0$ if $a^* > 0$.

The expected payoff to agent 1 if she chooses to rate when all other agents use the threshold strategy $a^*$ is
$$u_1^R = (1 - F(a^*))^{n-1}(p_R - c_R),$$
while the payoff to agents $i \geq 2$ from rating, when all remaining agents play according to their strategies $a^*$, is $u_i^R = p_R - c_R$. Note that $u_i^R \geq u_1^R$ for $i \geq 2$: this is because agent 1's strategy is to always contribute, which means that there is always content for the remaining agents to rate.

Now consider payoffs from contributing. If the system can perfectly distinguish the qualities of the contributions, then the probability that an agent $i$ finishes ahead of some particular other agent when she draws a quality equal to her threshold ($a_i = a_i^*$) is $F(a^*)$: if the other agent is $j \geq 2$, $i$ wins only if the other agent does not participate, which has probability $F(a^*)$; if the other agent is $j = 1$, $i$ wins if $a_1 \leq a_i^* = a^*$, which also has probability $F(a^*)$. If the system cannot distinguish qualities, and orders the contributions randomly, then the probability that an agent who contributes finishes ahead of some particular other agent $i \geq 2$ is $F(a^*) + \frac{1}{2}(1 - F(a^*)) = \frac{1}{2} + \frac{1}{2}F(a^*)$, while the probability that an agent who contributes finishes ahead of agent 1 is $\frac{1}{2} < \frac{1}{2} + \frac{1}{2}F(a^*)$ for $a^* > 0$.

Therefore, the distribution of the number of other agents that agent 1 beats, when she plays according to this strategy and has ability equal to her threshold, strictly first order stochastically dominates the distribution of the number of other agents than an agent $i \geq 2$ beats when contributing with ability equal to her threshold. This means that the expected payoff to agent 1 from contributing at *her* threshold, $u_1^C(a_1^*)$, is greater than the expected payoff $u_i^C(a^*)$ to other agents $i \geq 2$ from contributing at their threshold abilities.

Combining the facts that $u_i^R \geq u_1^R$ and $u_1^C(a_1^*) > u_i^C(a^*)$, we see that if an agent $i \geq 2$ is indifferent between rating and contributing when her ability $a_i$ is equal to her threshold $a^*$, then agent 1 strictly prefers contributing to rating when she draws ability $a_1 = a_1^*$. So if there is a value of $a^* > 0$ such that an agent $i \geq 2$ is indifferent between rating and contributing at ability $a_i = a^*$ (given that the remaining agents play according to their thresholds $a_1^* = 0$ and $a_i^* = a^*$), we will have
$$u_1^C(a_1^*) > u_i^C(a^*) = u_i^R \geq u_1^R,$$
in which case the threshold strategies $a_1^* = 0$ and $a_i^* = a^*$ constitute an (asymmetric) equilibrium.

To prove such a $a^*$ exists, note that we may assume that $a_1^* = 0$ and $a_i^* = a^* = 1$ is not an equilibrium, since otherwise we have already produced an asymmetric equilibrium as claimed. So suppose that it is not an equilibrium for agents $i \geq 2$ to use the threshold $a_i^* = 1$ and agent 1 to use $a_1^* = 0$. We will show that a solution to $u_i^C(a^*) = u_i^R$ must exist for some $a^* > 0$.

For $a^* = 1$, note that $u_i^C(a^*) > u_i^R$. To see this, suppose by means of contradiction that $u_i^C(a^*) \leq u_i^R$. Then none of the agents $i = 2, \ldots, n$ have any incentive to deviate under the thresholds $a_1^* = 0$ and $a_i^* = 1$. Since we assumed that these strategies do not constitute an equilibrium and agents $2, \ldots, n$ have no incentive to deviate, this means that agent 1 prefers to deviate from contributing to rating when all other agents are rating. Thus $p_1 - c_C \leq 0$, since the expected payoff from rating when no other agent is contributing is 0. But when $p_1 - c_C \leq 0$, it is an equilibrium for all agents to rate, contradicting the assumption in the statement of the theorem. So $u_i^C(q^*) > u_i^R$ must hold when $a^* = 1$.

Next note that if agents $i \geq 2$ use the threshold $a_i^* = 0$, so that all agents always contribute, then an agent with ability equal to the threshold must strictly prefer to deviate by rating because we have assumed that it is not an equilibrium for all agents to contribute content. Therefore, $u_i^C(a^*) < u_i^R$ at $a^* = 0$.

Thus if $w_\Delta(a^*)$ denotes the expected utility difference that an agent $j \geq 2$ obtains from rating rather than contributing when $j$ has ability $a_j = a^*$ and all remaining agents are playing according to their threshold strategies ($a_1^* = 0$ and $a_i^* = a^*$), then $w_\Delta(a^*) > 0$ when $a^* = 0$ and $w_\Delta(a^*) < 0$ when $a^* = 1$. But $w_\Delta(a^*)$ is continuous in $a^*$, so by the intermediate value theorem, there exists some $a^* \in (0, 1)$ such that $w_\Delta(a^*) = 0$. For this $a^*$, it is an equilibrium for agent 1 to always contribute and agents $i \geq 2$ to use the threshold strategy $a^*$. Thus an asymmetric equilibrium exists. □

This result indicates that while the mechanism designer can choose the rewards $p_B$ and $p_C$ in $\mathcal{M}_B$ so that there is an equilibrium in which $E[V]$ is maximized, it is not possible to ensure this is the unique equilibrium when allowing for asymmetric equilibria. However, this negative result also holds for a more general class of mechanisms: using more general rank-order mechanisms will not help to strengthen the implementation result.

We note that the class of mechanisms for which we explicitly constructed asymmetric equilibria is not the only class of mechanisms that is available to a mechanism designer, and is in fact a strict subset of the class of mechanisms for which we proved the optimal implementation result. We leave open the question of whether there is some alternative mechanism that would result in a unique equilibrium that implements the optimal outcome. Nonetheless, this result indicates that asymmetric equilibria remain a distinct possibility under a broad class of mechanisms corresponding to those most commonly used in practice.

## 4  Endogenous Effort

The implementability result in Theorem 3.3 was for the homogeneous effort model where all agents incur the same cost $c_C$ for contributing, as might be the case in expertise-based Q&A forums where the cost to answering is primarily the effort of transcribing one's knowledge. However, there are other settings where a contributor might be able to significantly increase the quality of her contribution by putting in more effort, as in research or effort-intensive questions, such as those on StackOverflow which request code to accomplish a task. Suppose that in addition to making strategic decisions about whether or not to contribute, an agent can also make strategic decisions about how much effort to exert on her contribution. In this section, we investigate the question of whether contests $\mathcal{M}_B(p_B, p_C)$ are still adequate to implement optimal outcomes in such an endogenous effort model where agents can strategically choose their effort or cost.

As before, to address the question of implementability, we must first answer the question of whether an equilibrium exists, and understand the form of equilibria. We begin with the following result.

**Theorem 4.1.** *There exists an equilibrium in which all agents use a symmetric threshold* participation *strategy where an agent participates if and only if her ability is greater than a common threshold $a^*$, and conditional on participating, each agent chooses an effort level using a symmetric strategy that is a function of her ability $a_i$.*

*Proof.* First note that each agent's expected payoff from participating is nondecreasing in $a$ because for any given level of effort $e$, the quality of the agent's contribution only increases with an increase in her ability $a$, and so her expected payoff from participating is at least as large as when participating with the lower ability. From this it follows that, regardless of the strategies the other agents are using, each agent has a best response which involves a threshold $a^*$ such that the agent participates if and only if the agent's ability $a \geq a^*$.

Now restrict attention to the set of strategies in which each agent $i$ chooses a cutoff $a_i^*$ such that the agent contributes if and only if $a_i \geq a_i^*$ and chooses effort strategies $e_i(a)$ for each $a$ such that she contributes according to the (possibly random) effort given by $e_i(a)$ if she draws ability $a$. Note that each individual's strategy space is compact: each possible choice variable for the agent is a number in the interval $[0, 1]$, which is compact, so we know from Tychonoff's Theorem that the strategy space, a product of compact spaces, is compact. The strategy space is also Hausdorff. Also note that the strategies and payoffs in this game are symmetric and each player's payoff is continuous in the strategies of the other players. Thus from [4], it follows that there exists a symmetric mixed strategy equilibrium to this game, and the result follows. □

Next we address the nature of the strategies that will maximize the mechanism designer's utility. As before, we restrict attention to symmetric strategies and discuss which subset of these symmetric strategies optimize the mechanism designer's utility. Again, we find that $V$ is maximized when only agents with abilities above a certain threshold participate; in addition, every such participating agent must exert the maximum possible effort.

**Theorem 4.2.** *There exists some $\hat{a} \in [0, 1]$ such that the mechanism designer's utility is maximized when an agent $i$ participates if and only if $a_i \geq \hat{a}$, and each participating agent chooses effort $e_i = 1$.*

*Proof.* First note that for any fixed participation strategies of the agents, the mechanism designer's utility is maximized when all agents choose effort $e_i = 1$ conditional on participating because choosing effort $e_i = 1$ results in higher quality contributions without changing the distribution of the number of agents who contribute. Thus the strategy that maximizes the mechanism designer's utility is contained in the subset of strategies in which the agents always exert effort $e_i = 1$. But we know from the reasoning in the proof of Lemma 3.2 that when the agents do not vary their levels of effort, then there exists some $\hat{a} \in [0,1]$ such that the mechanism designer's utility is maximized if each agent $i$ participates if and only if $a_i \geq \hat{a}$. Combining these arguments gives the result. □

We are now ready to address the question of whether outcomes that maximize the mechanism designer's utility can be supported in an equilibrium of the best contribution mechanism $\mathcal{M}_B(p_B, p_C)$ for any values of $p_B$ and $p_C$. Recall that when agents have homogeneous costs $c_C$, *i.e.*, they cannot endogenously choose their efforts to influence their contribution qualities, it is always possible to choose $p_B$ and $p_C$ to implement an optimal outcome. Our first result shows that in contrast, the best contribution mechanism need not be adequately powerful to incentivize optimal strategies if agents can endogenously choose their costs or levels of effort— surprisingly, this inability arises from perfectly ranking contributions' qualities.

**Theorem 4.3.** *Suppose that the mechanism $\mathcal{M}_B(p_B, p_C)$ always ranks contributions perfectly, i.e.,*

$$\pi(q_i, q_{-i}, m) = 1 \Leftrightarrow q_i > q_j$$

*for all contributing agents $j \neq i$. Then there is no equilibrium in which agents follow strategies that maximize the mechanism designer's utility $V$ for all values of $p_B$ and $p_C$.*

*Proof.* Suppose that all other agents are following a strategy that maximizes the mechanism designer's utility, and consider an agent $i$ with ability $a_i = \hat{a}$, where $\hat{a}$ denotes the threshold according to which agents contribute in this strategy. Note that such an agent will be selected as the winner if and only if she is the only agent who contributes, because any other agent $j$ who contributes content has greater ability and exerts effort $e_j = 1$ and therefore has a higher quality. But this means that agent $i$'s choice of effort has no effect on the agent's probability of being selected as the winner. So if this agent contributes, then it is a best response for her to exert effort $e_i = 0$.

Note that this low effort best response is not an artifact of choosing an agent with ability exactly equal to the threshold: a similar argument shows that if an agent has ability $a_i = \hat{a} + \epsilon$ for some small $\epsilon > 0$, then her best response cannot be to choose $e_i = 1$. From this it follows that there is no equilibrium in which all agents with abilities $a_i \in [\hat{a}, 1]$ choose $e_i = 1$, *i.e.*, follow strategies that maximize the mechanism designer's utility. □

Thus if the system can (and chooses to) rank contributions perfectly in decreasing order of quality, it will fail to create incentives for agents to exert optimal levels of effort in equilibrium. The intuitive reason behind this result is that when a mechanism ranks contributions perfectly, a contributor with low ability $a_i$ knows that she has no chance of being ranked ahead of higher ability contributors (when they also exert high effort) regardless of how much effort she exerts, since $q(a, e)$ is increasing in both $a$ and $e$. So she has no incentive to expend effort to try to improve the quality of her contribution, and it is not an equilibrium for contributors to exert the maximal level of effort as required to optimize the mechanism designer's utility.

However, our next result shows that while perfect rankings of the contributions always create incentives for agents to avoid exerting the optimal level of effort, these adverse incentives can be avoided if the system does not rank the contributions perfectly in order of their qualities, for example by perturbing the rankings by adding random noise to $q(a_i, e_i)$. With such noisy rankings, exerting extra effort can always increase the probability that a contributor will ultimately be (possibly erroneously) ranked ahead of other participants and be selected as the best contribution, giving agents a greater incentive to exert effort than when qualities of contributions are measured perfectly. In particular, the mechanism designer may then be able to induce agents to follow strategies that maximize the mechanism designer's utility if the cost of effort $c(e)$ does not grow too rapidly, as the following theorem illustrates.

**Theorem 4.4.** *Consider any $p_R \geq 0$ and any particular ratio of the rewards $p_B/p_C > 1$. Suppose that*
$$c'(e_i) \leq \frac{\partial q_i}{\partial e_i} \frac{\partial \pi}{\partial q_i} (p_B - p_C)(1 - F(\hat{a})^{n-1})$$
*for all $e_i, a_i, q_{-i}$ and $m \geq 2$ when $p_B = c(0)$ and $\hat{a}$ denotes the optimal threshold in equilibrium. Then the mechanism designer can choose the values of $p_B$ and $p_C$ such that it is an equilibrium for agents to follow strategies that maximize the mechanism designer's utility.*

*Proof.* Consider agent $i$, and suppose at least one other agent makes a contribution. The benefit $b_i(q_i, q_{-i})$ to $i$ from contributing is
$$b_i(q_i, q_{-i}) = \pi(q_i, q_{-i}) p_B + (1 - \pi(q_i, q_{-i})) p_C.$$
So the marginal benefit to agent $i$ from exerting greater effort when at least one other agent contributes is
$$\frac{\partial b_i}{\partial e_i} = \frac{\partial q_i}{\partial e_i} \frac{\partial \pi}{\partial q_i} (p_B - p_C).$$
Now if agents follow strategies in which they contribute if and only if $a_i \geq a^*$ for some $a^* \in [0, \hat{a}]$, then the probability at least one other agent contributes is greater than or equal to $1 - F(\hat{a})^{n-1}$. Thus the expected marginal benefit to agent $i$ from exerting greater effort if all other agents follow strategies in which they contribute if and only if $a_i \geq a^*$ for some $a^* \in [0, \hat{a}]$ is at least $\frac{\partial q_i}{\partial e_i} \frac{\partial \pi}{\partial q_i} (p_B - p_C)(1 - F(\hat{a})^{n-1})$.

Now if $p_B \geq c(0)$, then for any fixed $p_B/p_C > 1$, $p_B - p_C$ is minimized when $p_B$ assumes the minimum possible value in this range, $c(0)$. So if
$$c'(e_i) \leq \frac{\partial q_i}{\partial e_i} \frac{\partial \pi}{\partial q_i} (p_B - p_C)(1 - F(\hat{a})^{n-1})$$
when $p_B = c(0)$, then $c'(e_i) \leq \frac{\partial q_i}{\partial e_i} \frac{\partial \pi}{\partial q_i} (p_B - p_C)(1 - F(\hat{a})^{n-1})$ for all $p_B \geq c(0)$.

Combining these arguments shows that if $p_B \geq c(0)$ and all agents follow strategies in which they contribute if and only if $a_i \geq a^*$ for some $a^* \in [0, \hat{a}]$, then the expected marginal benefit to agent $i$ from exerting greater effort is greater than or equal to the value of $\frac{\partial q_i}{\partial e_i} \frac{\partial \pi}{\partial q_i} (p_B - p_C)(1 - F(\hat{a})^{n-1})$ when $p_B = c(0)$. Thus if $c'(e_i) \leq \frac{\partial q_i}{\partial e_i} \frac{\partial \pi}{\partial q_i} (p_B - p_C)(1 - F(\hat{a})^{n-1})$ when $p_B = c(0)$, then the marginal costs from exerting greater effort are no greater than the expected marginal benefits from exerting greater effort if $p_B \geq c(0)$ and all agents follow strategies in which they contribute if and only if $a_i \geq a^*$ for some $a^* \in [0, \hat{a}]$. From this it follows that if $p_B \geq c(0)$ and all agents follow strategies in which they contribute if and only if $a_i \geq a^*$ for some $a^* \in [0, \hat{a}]$, then any contributing agent has an incentive to exert effort $e_i = 1$.

Now consider some $p_B$ that is sufficiently large that all agents with abilities $a_i \geq a^*$ participate in equilibrium for some $a^* \in [0, \hat{a}]$. Note that this $p_B$ satisfies $p_B > c(0)$ because when $p_B = c(0)$, no agents participate in equilibrium. Thus for any such $p_B$ we know from the previous paragraph that all agents who contribute exert effort $e_i = 1$. Given that all agents are exerting effort $e_i = 1$, we also know from the same reasoning in Theorem 3.2 that the threshold $a^*$ that agents use in equilibrium varies continuously with $p_B$ and $p_C$.

Now let $p_B$ denote the smallest value of the points awarded for best answer such that it is an equilibrium for all agents with abilities $a_i \geq a^*$ to participate for some $a^* \in [0, \hat{a}]$. We claim that for such a $p_B$, it is an equilibrium for agents to participate if and only if $a_i \geq \hat{a}$. To see this, suppose by means of contradiction that this is not the equilibrium in this setting. Then we know that there exists some equilibrium in which agents participate if and only if $a_i \geq a^*$ for some $a^* < \hat{a}$ and exert effort $e_i = 1$ conditional on participating. But then from the reasoning used to show that the threshold $a^*$ that agents use in equilibrium varies continuously with $p_B$ and $p_C$, it follows that for slightly lower values of $p_B$, it is still an equilibrium for agents to participate if and only if $a_i \geq a^{**}$ for some $a^{**} < \hat{a}$ and exert effort $e_i = 1$ conditional on participating. This contradicts the fact that $p_B$ is the smallest value of the points awarded for best answer such that it is an equilibrium for all agents with abilities $a_i \geq a^*$ to participate in equilibrium for some $a^* \in [0, \hat{a}]$, and proves that for such a $p_B$, it is an equilibrium for agents to participate if and only if $a_i \geq \hat{a}$.

But this indicates that for any fixed ratio of the rewards $p_B/p_C$, there exists some $p_B$ such that it is an equilibrium for agents to participate if and only if $a_i \geq \hat{a}$ and exert effort $e_i = 1$ conditional on participating. The result then follows. □

*Perfect and noisy rankings.* If contributions are ranked perfectly according to their qualities, then $\frac{\partial \pi}{\partial q_i} = 0$ for almost all values of $q$ because incremental changes in $q_i$ do not affect the fact that the highest quality contribution is always chosen as the best. Similarly, if the winning contribution is chosen completely randomly without regard to the qualities of the contributions, then $\frac{\partial \pi}{\partial q_i} = 0$ also holds because qualities have no affect on the rankings. However, if we rank contributions according to a noisy signal of quality $s_i = q_i + \epsilon_i$, where $\epsilon_i$ is drawn from some distribution with adequately large support, then $\frac{\partial \pi}{\partial q_i} > 0$ because incremental improvements in $q_i$ always increase a contribution's probability of being selected as the best answer.

Theorem 4.4 says that if the marginal cost of exerting effort, $c'(e)$, is not too large and the system does not rank the qualities of the contributions perfectly, so that $\frac{\partial \pi}{\partial q_i} > 0$ always holds when there are at least two contributions, then the values of $p_B$ and $p_C$ can be chosen so that agents follow strategies that maximize the mechanism designer's utility in an equilibrium of $\mathcal{M}_B(p_B, p_C)$. We note that the condition on the cost function in this result can be weakened to only requiring the inequality in theorem statement to hold for $p_B = c(0) + \max\{p_R - c_R, 0\}$; we choose the current statement for simplicity and to be obviously applicable irrespective of whether the system allows rating or not.

Together, Theorems 4.3 and 4.4 suggest that a system that does not rank contributions perfectly (of course, the ranking must still remain monotone in quality) may induce better outcomes than a system that does perfectly rank contributions, because such perturbed rankings can create incentives for agents to exert greater effort. While our results require a condition that the cost function does not grow too rapidly, and therefore do not necessarily imply this will hold under full generality, they do indicate that the designer may wish to intentionally induce noise in these rankings when the marginal cost of effort is not too large to allow implementation of optimal outcomes.

We note that this reasoning, indicating that it can be better for a mechanism designer not to have precise information about the qualities of contributions, parallels a result in [2] showing that a firm may find it beneficial not to perfectly monitor its workers if it wants to create incentives for its employees to work hard. [2] finds that with perfect monitoring, the higher ability workers will always do better than the lower ability workers, and employees will have no incentive to work harder than necessary to maintain their position. However, if a firm does not monitor its workers perfectly, then employees have an incentive to work hard because extra effort can always increase the probability that they will ultimately be (possibly erroneously) ranked ahead of their coworkers.

## 5  Discussion

In this paper, we analyzed the effectiveness of the widely used best contribution mechanism at implementing optimal outcomes, *i.e.*, at inducing selfish agents to choose strategies that achieve the maximum utility the mechanism designer could obtain if agents were nonstrategic. We conclude by discussing the robustness of our results to the assumption that users are homogeneous in their motivations to contribute. While most users may be well-described by the strategic incentives in our model, there may also be users who contribute *non-strategically*, without regard to any gain of points (as an example, there might be heavily invested users who always supply high quality answers for topics they care about, without regard to point-based incentives). Such contributors can be modeled as nonstrategic agents whose participation choices and qualities are exogenous. Our results about when optimal implementation can be achieved all continue to hold in this case, using similar arguments. For instance, in the case of homogenous effort, strategic agents would continue to follow symmetric threshold strategies in equilibrium (the specific thresholds would depend on the distribution of the qualities of the non-strategic agents), and the equilibrium threshold would again vary continuously with the rewards $p_B$ and $p_C$. The mechanism designer would continue to want the strategic agents to use some symmetric threshold strategy in equilibrium (again, the value of the desired threshold might depend on the distribution of the nonstrategic agents' qualities). The same argument as in Theorem 3.3 would then show that the mechanism designer could achieve optimal outcomes by choosing rewards appropriately.

**Acknowledgments.** We thank Preston McAfee and David Pennock for helpful discussions.

# A  Comparative Statics and Learning

In this section, we outline how the mechanism designer would want to set the values of the rewards $p_B$ and $p_C$ so that it is an equilibrium for agents to follow strategies that maximize the mechanism designer's utility. Throughout this section, we focus on the *homogenous effort model*.

**Comparative Statics**  First we discuss comparative statics, namely how the values of $p_B$ and $p_C$ that must be used to induce the optimal outcome vary as a function of the number of agents, $n$. Such results are relevant because some types of questions differ systematically in the number of agents that arrive: for example, popular topics of general interest in Yahoo! Answers, such as entertainment, typically attract a lot of agents, while more specialized topics, such as theoretical physics, attract fewer agents. While there are no general comparative statics results when the mechanism designer's utility $V$ can vary with the number of contributions in an arbitrary way, as we have allowed for in all our results so far, comparative statics results can be obtained for natural restrictions in the mechanism designer's utility. Here we consider what happens when $\hat{a}(n)$, the ideal threshold that the mechanism designer would like the agents to use, is independent of $n$. This setting arises naturally, for instance, if the mechanism designer would like all agents with adequately high abilities to contribute, but not the remaining agents. We obtain the following result.

**Proposition A.1.** *Suppose $\hat{a}(n) \in (0,1)$ is independent of $n$. For any fixed $p_R$ and any fixed ratio $p_B/p_C > 1$, the numbers of points the mechanism designer should award as a function of $n$, $p_B(n)$ and $p_C(n)$, are strictly increasing in $n$.*

*Proof.* To induce agents to use a threshold strategy $\hat{a}(n) \in (0,1)$, the values of $p_B$ and $p_C$ must be such that $u_\Delta(p_B, p_C, a, n) \equiv (p_R - c_R)Pr(\mathcal{C}_{>0}|a,n) - (p_B - c_C)Pr(\mathcal{W}|a,n) - (p_C - c_C)[1 - Pr(\mathcal{W}|a,n)] = 0$ when $a = \hat{a}(n)$. Now $u_\Delta(p_B, p_C, a, n)$ is strictly decreasing in $p_B$ and $p_C$ when $a = \hat{a}(n) \in (0,1)$. Also, $u_\Delta(p_B, p_C, a, n)$ is strictly increasing in $n$ for any fixed $a = \hat{a}(n) \in (0,1)$ because $Pr(\mathcal{C}_{>0}|a,n)$ is strictly increasing in $n$ and $Pr(\mathcal{W}|a,n)$ is strictly decreasing in $n$ for any fixed $a = \hat{a}(n) \in (0,1)$. So the numbers of points the mechanism designer should award for contributions as a function of $n$, $p_B(n)$ and $p_C(n)$, are strictly increasing in $n$. □

We note that the result in Proposition A.1 holds under more general assumptions than that $\hat{a}(n)$ is independent of $n$. As long as $\hat{a}(n)$ is nonincreasing in $n$, and the mechanism designer wants to induce at least as large a rate of participation when there are more agents, then he will still want to award points in such a way that agents will have a greater incentive to contribute when there are a larger number of agents. Thus the result in Theorem A.1 continues to hold if $\hat{a}(n)$ is nonincreasing in $n$.

**Learning the Parameter Values**  We have assumed so far that the mechanism designer knows the agents' costs for contributing and rating, $c_C$ and $c_R$, which are used to determine the rewards $p_B$ and $p_C$ that induce agents to use optimal strategies. While it is reasonable to assume that agents know their own preferences over actions, and in particular that agents know how costly it is for them to contribute or rate, it may not always be reasonable to assume that the mechanism designer knows these costs as well. In this section, we illustrate how a mechanism designer may learn the agents' costs by observing their behavior.

We first consider a scenario in which the mechanism designer knows the value of $c_R$ but does not know the precise value of $c_C$. The assumption that the mechanism designer knows the value of $c_R$ is reasonable in many settings because there are many contexts in which rating content takes almost no effort at all, i.e., $c_R = 0$. We will later illustrate how values may be learned when the mechanism designer also does not know the value of $c_R$.

We assume throughout that the mechanism designer only knows that $c_C \in (0, \overline{c})$ for some $\overline{c} \in (0, \infty)$; thus there is some upper bound on the maximum possible cost an agent could incur from contributing content. The agents know how costly it is for them to contribute content, so if the mechanism designer awards $p_B$ points for winning, $p_C$ points for contributing but not winning, and $p_R$ points for rating, then the agents play according to the same equilibrium strategies given in §3.1. In particular, agents use a threshold $a^*(c_C)$, which is the unique symmetric equilibrium that agents can use when their cost for contributing is $c_C$.

Suppose the mechanism designer conducts a series of $T$ contests, where he awards $p_B$ points for winning, $p_C$ points for contributing but not winning, and $p_R$ points for rating in each contest. We let $n_C^t$ denote the number of agents who contribute content in the $t^{th}$ contest and suppose that the number of agents $n$ in each contest is the same for all contests $t$. We have the following easy lemma.

**Lemma A.1.** *If the mechanism designer conducts a series of $T$ contests with $n$ agents in which the mechanism designer awards the same numbers of points in all contests, then $\sum_{t=1}^{T} n_C^t/(Tn)$ converges in probability to $1 - F(a^*(c_C))$ as $T \to \infty$.*

*Proof.* By the law of large numbers, $\sum_{t=1}^{T} n_C^t/(Tn)$ converges in probability to the probability that an agent contributes content in equilibrium as $T \to \infty$. This probability is $1 - F(a^*(c_C))$. □

This result guarantees that the mechanism designer can learn the frequency with which agents contribute content, $1 - F(a^*(c_C))$, by conducting a large number of contests in which the mechanism designer awards the same numbers of points in each contest. From this the mechanism designer can immediately learn the threshold $a^*(c_C)$ that agents are using in equilibrium when there are $p_B$ points for winning, $p_C$ points for contributing but not winning, and $p_R$ points for rating. This in turn implies that if there is a one-to-one correspondence between values of $c_C$ and values of $a^*(c_C)$, then the mechanism designer can learn the value of $c_C$ by conducting the series of contests described above. We now address when there is a one-to-one correspondence between values of $c_C$ and values of $a^*(c_C)$.

**Lemma A.2.** $a^*(c_C)$ *is strictly increasing in $c_C$ for values of $c_C$ satisfying $a^*(c_C) \in (0, 1)$.*

*Proof.* Note that if $a^*(c_C) \in (0, 1)$, then $a^*(c_C)$ is the value of $a$ that satisfies the equation $u_\Delta(c_C, a) \equiv (p_R - c_R)Pr(\mathcal{C}_{>0}|a) - (p_B - c_C)Pr(\mathcal{W}|a) - (p_C - c_C)[1 - Pr(\mathcal{W}|a)] = 0$. Now we have seen that $u_\Delta$ is strictly decreasing in $a$. Also, from the above expression, it follows that $u_\Delta(c_C, a)$ is strictly increasing in $c_C$. From this it follows that the value of $a$ that satisfies the equation $u_\Delta(c_C, a) = 0$ is strictly increasing in $c_C$ for values of $c_C$ satisfying $a^*(c_C) \in (0, 1)$. □

By combining these two lemmas, we obtain the main result of this section.

**Theorem A.1.** *If the mechanism designer conducts a series of $T$ contests with $n$ agents in which the mechanism designer awards $p_B > p_R + \overline{c}$ points for winning and $p_C = 0$ points for contributing but not winning, then $\sum_{t=1}^{T} n_C^t/(Tn)$ converges in probability to an invertible function of $c_C$ as $T \to \infty$.*

*Proof.* From the previous two lemmas, we know that if $a^*(c_C) \in (0,1)$ always holds, then $\sum_{t=1}^{T} n_C^t/(Tn)$ converges in probability to an invertible function of $c_C$ as $T \to \infty$. But if $p_C = 0$ and $p_B > p_R + \overline{c}$, then $p_B - c_C > p_R - c_R > p_C - c_C$ necessarily holds. We have already seen that if $p_B - c_C > p_R - c_R > p_C - c_C$, then the agents play according to a symmetric threshold equilibrium with threshold $a^* \in (0,1)$. Thus if $p_C = 0$ and $p_B > p_R + \overline{c}$, then $\sum_{t=1}^{T} n_C^t/(Tn)$ converges in probability to an invertible function of $c_C$ as $T \to \infty$. $\square$

Thus the mechanism designer can use the series of contests described above to infer how costly it is for agents to contribute content with arbitrary precision in the limit of a large number of contests, and deduce the optimal levels of points necessary to induce the agents to follow strategies that maximize his utility. We note that it is not necessary to observe how agents react to a large number of different rewards in order to learn the optimal values of $p_B$ and $p_C$ for a given population. The mechanism designer can simply award a constant level of points for a long time, observe how agents react to this, and then use this information to deduce the optimal values for $p_B$ and $p_C$. Thus this method of learning the optimal rewards is one that can be easily implemented in practice.

The learning result in this section has made use of the assumption that the mechanism designer knows $c_R$. When $c_R$ is not known, there may be multiple values of $c_C$ and $c_R$ that would result in the agents using a given threshold for a particular value of $p_B$. However, the mechanism designer could learn the values of $c_C$ and $c_R$ by conducting an additional experiment similar to that in Theorem A.1, but with a larger number of points for winning, *i.e.*, a larger value of $p_B$. The agents will use different thresholds in these two different contests, and the mechanism designer could use the learned values of these thresholds to obtain two non-degenerate linear equations for the unknowns $c_C$ and $c_R$ that uniquely determine the values of these costs. Thus if $c_C$ and $c_R$ are not known, the mechanism designer can learn the values of these parameters by conducting two separate experiments of the form in Theorem A.1 with different levels of points for winning.